\documentclass[aps,prb,twocolumn,showpacs,groupedaddress]{revtex4-1}
\usepackage{amssymb}
\usepackage{hyperref}
\usepackage{graphicx}
\usepackage{amsmath}

\begin{document}

\title[]{Quantum oscillations in metallic Sb$_2$Te$_2$Se topological insulator}
\author{K. Shrestha$^{1}$}
\email[E-mail me at:]{keshav.shrestha@inl.gov}
\author{ V. Marinova$^{2}$, D. Graf$^{3}$, B. Lorenz$^{4}$ and C. W. Chu$^{4,}$$^{5}$}
\affiliation{$^{1}$Idaho National Laboratory,2525 Fremont Ave, Idaho Falls, ID 83401, USA}
\affiliation{$^{2}$Institute of Optical Materials and Technology, Bulgarian Academy of Sciences, Acad. G. Bontchev Street 109, Sofia 1113, Bulgaria}
\affiliation{$^{3}$National High Magnetic Field Laboratory, Florida State University, Tallahassee, Florida 32306-4005, USA}
\affiliation{$^{4}$TCSUH and Department of Physics, University of Houston, 3201 Cullen Boulevard, Houston, Texas 77204, USA}
\affiliation{$^{5}$Lawrence Berkeley National Laboratory, 1 Cyclotron Road, Berkeley, California 94720, USA}

\begin{abstract}
  We have studied the magnetotransport properties of the metallic, $p$-type Sb$_2$Te$_2$Se which is a topological insulator. Magnetoresistance shows Shubnikov de Haas oscillations in  fields above $B$=15 T. The maxima/minima positions of oscillations measured at different tilt angles with respect to the $B$ direction align with the normal component of field $B$cos$\theta$, implying the existence of a 2D Fermi surface in Sb$_2$Te$_2$Se. The value of the Berry phase $\beta$ determined from a Landau level fan diagram is very close to 0.5, further suggesting that the oscillations result from topological surface states. From Lifshitz-Kosevich analyses, the position of the Fermi level is found to be $E_F$=250 meV, above the Dirac point. This value of $E_F$ is almost 3 times as large as that in our previous study on the Bi$_2$Se$_{2.1}$Te$_{0.9}$ topological insulator; however, it still touches the tip of the bulk valence band. This explains the metallic behavior and hole-like bulk charge carriers in the Sb$_2$Te$_2$Se compound.

\end{abstract}


\maketitle

\indent Novel quantum phenomena arising from the topologically protected surface state in systems with strong spin-orbit coupling have been a focus of interest for several years \cite{Hasan:01}$^{,}$\cite{Qi:02}$^{,}$\cite{Ando:03}$^{,}$\cite{Ando1:04}$^{,}$\cite{Cava:05}. In topological insulators, the nontrivial topology of the bulk band structure results in conductive electronic states at the surface since the transition to the trivial topology of the vacuum demands  passing through gapless boundary states. The gapless surface states are topologically protected and less sensitive to perturbations or impurities, and their excitation spectrum within the bulk energy gap exhibits the characteristic Dirac dispersion.\\
\indent Bulk and surface states can be observed through angle resolved photoelectron spectroscopy\cite{Hasan:01}$^{,}$\cite{Qi:02}$^{,}$\cite{Ando:03}. Electrical transport measurements in magnetic fields are frequently used to gather information about the details of the Fermi surface of bulk and/or surface states. Quantum oscillations of the conductivity (Shubnikov-de Haas, SdH) have their origin in the Landau level splitting of electron orbits in magnetic fields. For bulk carriers, the characteristic SdH oscillation frequencies provide information about the Fermi surface and their dependence on the orientation of the magnetic field probes the shape of the Fermi surface. Surface carriers, however, are bound to the two-dimensional character of the surface and therefore the SdH oscillation frequencies depend on the field component normal to the surface. Furthermore, the Dirac character of the surface carriers can be shown by evaluating the temperature and field dependence of the SdH oscillations and by extracting the Berry phase.\\
\indent In a perfect topological insulator, the bulk Fermi energy falls into a gap and the only contribution to the conductivity at sufficiently low temperature is from the topological surface states. Therefore, attention has been paid to the growth and synthesis of compounds with a minimum amount of defects or impurities\cite{Taskin:06}$^{,}$\cite{Gaku:07}. However, many topological insulators are not insulating in the bulk due to defects, impurities, or intersite defects. These effects usually result in a shift of the bulk Fermi energy into the valence (down) or conduction (up) bands and the material shows bulk conduction, as for example in the system Bi$_2$Se$_3$-Bi$_2$Te$_3$ \cite{Xia:08}$^,$\cite{Chen:09}$^,$\cite{Hsieh:10}. The bulk states contribute to the conduction in parallel to the surface transport and the interference of both conduction paths makes it more difficult, if not impossible, to detect the typical signature of the topological surface states in magnetotransport experiments\cite{Qu:11}$^,$\cite{Analytis:12}$^,$\cite{Eto:13}$^,$\cite{Cao:14}.\\
\indent In a recent study, however, we showed that SdH oscillations from bulk as well as surface carriers can be observed simultaneously in metallic Bi$_2$Se$_{2.1}$Te$_{0.9}$ \cite{Shrestha:15}$^,$\cite{Shrestha1:16}. Conductance oscillations from bulk and topological surface states appeared to be well separated in the high- and low-field regions, respectively. This result revealed the possibility of characterizing topological surface properties through magnetoconductance measurements even in materials with bulk metallic conduction. It is therefore of interest to extend the study of metallic topological systems and search for other candidates
where surface SdH oscillations can be resolved despite metallic conduction in the bulk.\\
\indent Sb$_2$Te$_2$Se is a member of a larger class of ternary tetradymite-like compounds \cite{Xu:17} and predicted to be a topological insulator \cite{Lin:18}$^,$\cite{Fu:20}. Wang $et$ $al$\cite{Wang:19} observed quantum oscillations in magnetoresistance and magnetic torque measurements in Sb$_2$Te$_2$Se under high fields up to 35 T. From the angle dependence of the quantum oscillations frequency, they reported an existence of a quasi-two dimensional Fermi surface originated from the bulk states. However, the Berry phase, which is the critical parameter to distinguish a topological material from non-topological materials, is lacking in their work. In this study, we have investigated the SdH oscillations in magnetoresistance of Sb$_2$Te$_2$Se. From the angle dependence of quantum oscillations and the determination of the Berry phase, we prove that the SdH oscillations are from charge carriers of topological surface states. No quantum oscillations from bulk carriers are detected up the maximum field of 31 T and the possible reasons are discussed.

High quality single crystals of Sb$_2$Te$_2$Se were grown by the modified Bridgman method. First, binary compounds Sb$_2$Se$_3$ and Sb$_2$Te$_3$ were synthesized. The synthesis was done by using stoichiometric quantities of the starting materials Sb, Se, and Te, each with purity of 99.9999\%, mixed in quartz ampoules with diameters of 20 mm and vacuum pumped to 10$^{-6}$ torr. The synthesis and homogenization process lasts for 25 hours at a temperature in the range of 620-650 $^{o}$C. The binary compositions prepared in this way were mixed to the desired ternary compounds, placed in quartz ampoules with diameters of 10 mm, vacuum pumped to 10$^{-6}$ torr, and sealed. These ampoules were further positioned in a Bridgman crystal growth furnace. In the furnace the ampoules were heated to 650 $^{o}$C and homogenized for 36 hours. The crystal growth process was performed through temperature decreasing with a speed of 0.5 $^o$C per hour in the range of 650-570 $^o$C. The ampoules were further cooled from 570 $^{o}$C to room temperature at a speed of 10 $^o$C per hour.\\
\indent Magnetotransport measurements were performed at the National High Magnetic Field Laboratory (NHMFL), with fields up to 31 T. Six gold contacts were sputtered on a freshly cleaved crystal face for standard resistivity and Hall measurements. Platinum wires were attached using silver paint. The sample was then mounted on the rotating platform of the standard probe designed at NHMFL. Longitudinal and Hall resistances were measured using a lock-in amplifier (SR 830). AC current of 1 mA was passed through the sample using a Keithley (6221) source meter. The longitudinal and Hall resistances were measured using two lock-in amplifiers, respectively. The sample was mounted on a rotating platform that can be positioned at different angles with respect to the applied magnetic field. The platform, mounted in a $^3$He Oxford cryostat, was inserted into the bore of a resistive magnet with a maximum field of 31 T. The position of the sample with respect to the applied field was calibrated by using a Hall sensor. Thermoelectric power was measured using an ac technique at low frequency (0.1 Hz). Two heaters generate a sinusoidal temperature gradient with amplitude of 0.25 K at a certain frequency, and the sinusoidal thermovoltage across the sample was measured at the same frequency.

\begin{figure}
  \centering
  \includegraphics[width=1.0\linewidth]{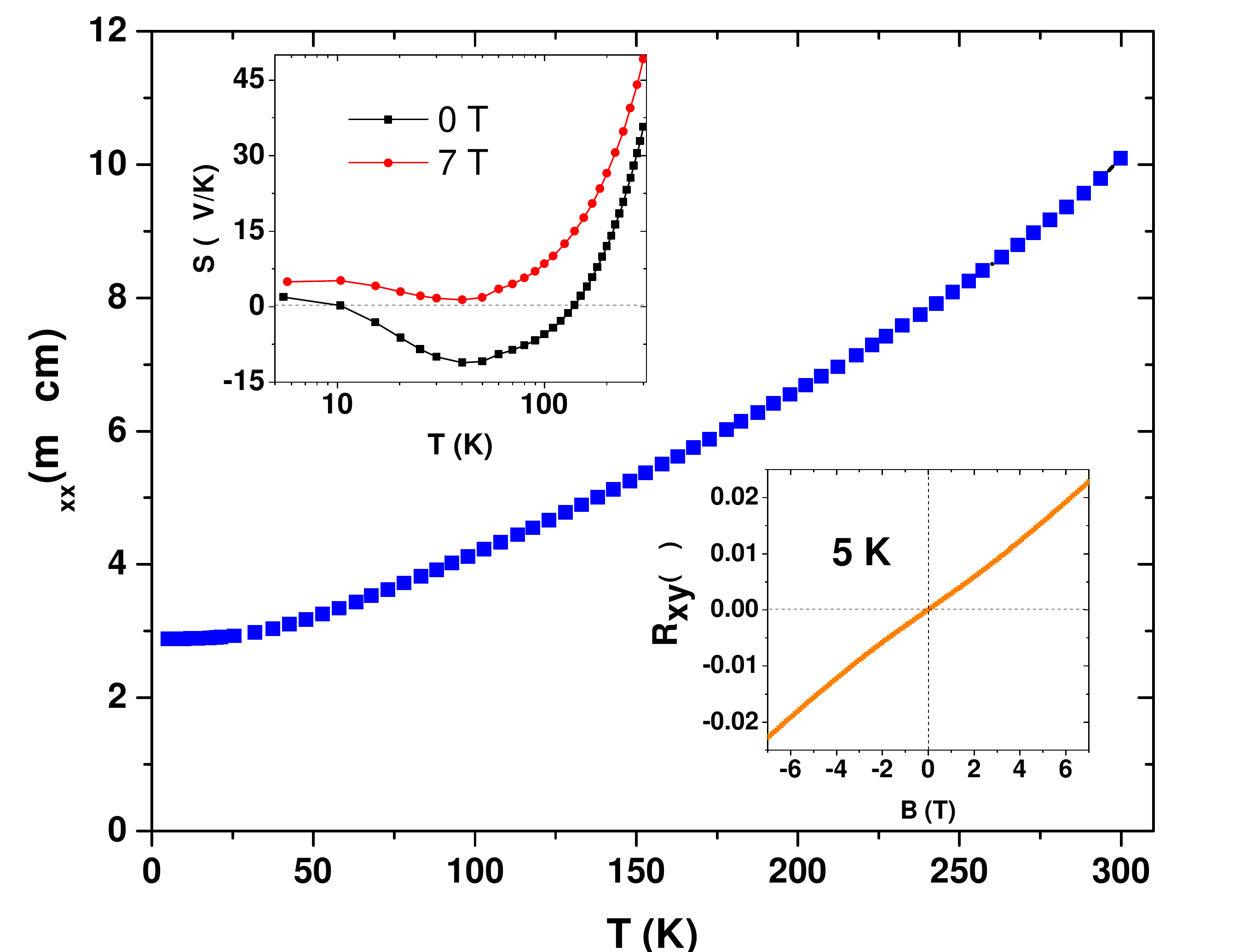}
  \caption{Temperature dependence of resistivity of a Sb$_2$Te$_2$Se single crystal. The lower right inset shows Hall data at 5 K. The upper left inset displays the logarithmic temperature dependence of the thermoelectric power at 0 (black) and 7 T (red) fields.}\label{Resistivity}
\end{figure}
Figure [1] shows the longitudinal resistivity as a function of temperature for a Sb$_2$Te$_2$Se single crystal. The sample exhibits a metallic behavior from 300 to 5 K. The residual resistivity ratio RRR=$\rho_{xx}$(300K)/$\rho_{xx}$(5K)=3.5 for the Sb$_2$Te$_2$Se single crystal indicates good crystalline quality. Hall measurements were carried out to determine the nature of the bulk charge carriers and their concentration. The positive slope of Hall resistance (lower right inset in figure [1]) confirms hole-like bulk charge carriers. Also, the bulk carrier concentration is estimated to be 3$\times$10$^{18}$cm$^{-3}$ at 5 K. Seebeck coefficient, $S$, as a function of logarithmic temperature is displayed in the upper left inset in figure [1]. Zero-field $S$ is positive above 150 K, becomes negative below 100 K, and re-enters the positive region below 10 K. This suggests that either electron- or hole-like charge carriers dominate at various temperature regions for Seebeck coefficient of Sb$_2$Te$_2$Se. At 5 K, the positive value of $S$ is consistent with the $p$-type nature of bulk charge carriers determined by the Hall measurements. With an application of 7 T field, $S$ is positive throughout the entire temperature range.

\begin{figure}
  \centering
  \includegraphics[width=1.0\linewidth]{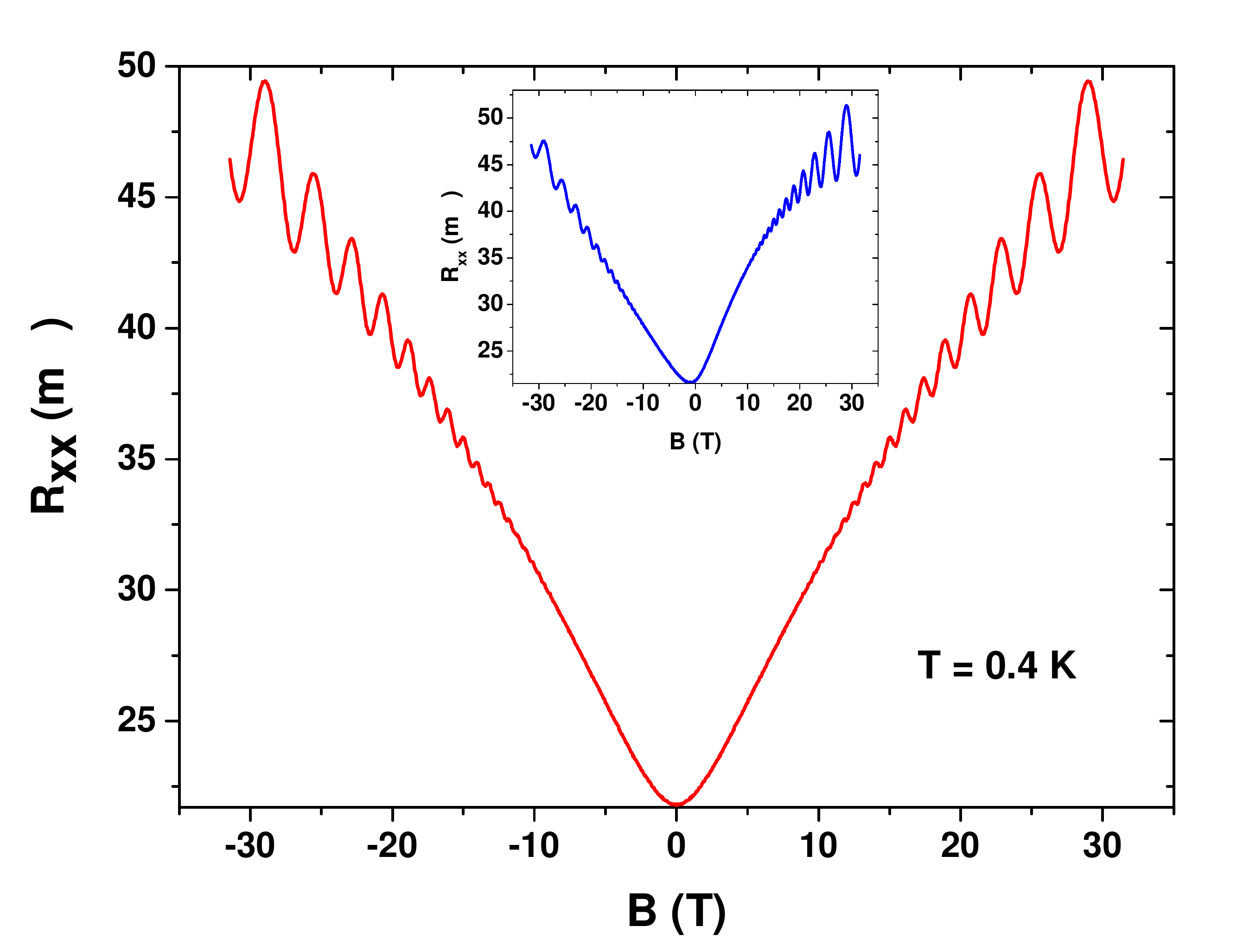}
  \caption{The symmetrized magnetoresistance, obtained by [$R_{xx}(B)$+$R_{xx}(-B)$]/2, of a Sb$_2$Te$_2$Se single crystal under high field up to 31 T at $T$=0.4 K. The raw magnetoresistance data of the Sb$_2$Te$_2$Se single crystal is shown in the inset.}\label{SdH}
\end{figure}
 The magnetoresistance of Sb$_2$Te$_2$Se is measured under high magnetic field up to 31 T at the NHMFL. The magnetoresistance is  anisotropic in the magnetic field direction as shown in the inset of figure [2]. However, the anisotropy can be removed by  taking the average of the magnetoresistance in the positive and negative field directions, i.e. [$R_{xx}(B)$+$R_{xx}(-B)$]/2. Figure [2] shows the symmetrized magnetoresistance at $T$=0.4 K with field along the c-axis. Magnetoresistance is positive and shows SdH oscillations above 15 T. Figure [3(a)] shows the quantum oscillations obtained after subtracting a smooth polynomial background at different temperatures. The oscillations are clear, and periodic in 1/$B$, showing the existence of a well-defined Fermi surface in Sb$_2$Te$_2$Se. Figure [3(b)] displays frequencies of quantum oscillations, obtained by taking a fast Fourier transform (FFT) of the quantum oscillations shown in figure [3(a)]. There exists a single frequency at $f$=215 T. The amplitude of the frequency decreases with an increase in temperature but the frequency remains the same for all temperatures.\\
\indent The angle dependence of quantum oscillations at different tilt angles $\theta$ provides the information about the shape, size, and dimensionality of the Fermi surface. Figure [4(a)] shows magnetoresistance at different $\theta$ values. The positions of maxima/minima change systematically with tilt angle $\theta$. The maxima/minima positions align with one another with the normal component of magnetic field, $B$cos$\theta$, as shown in figure [4(b)]. The oscillations depend only on the normal component of $B$, indicating the 2D nature of the Fermi surface. This provides a strong evidence that the observed quantum oscillations result from topological surface states\cite{Ren:21}. The amplitude of the quantum oscillations decreases at higher angle, and the oscillations could not be resolved in magnetorestance and its derivative for angles above 40$^{o}$. This further supports oscillations originating from surface states, i.e. Dirac fermions\cite{Ando:03}$^,$\cite{Qu:11}.
\begin{figure}
  \centering
  \includegraphics[width=1.0\linewidth]{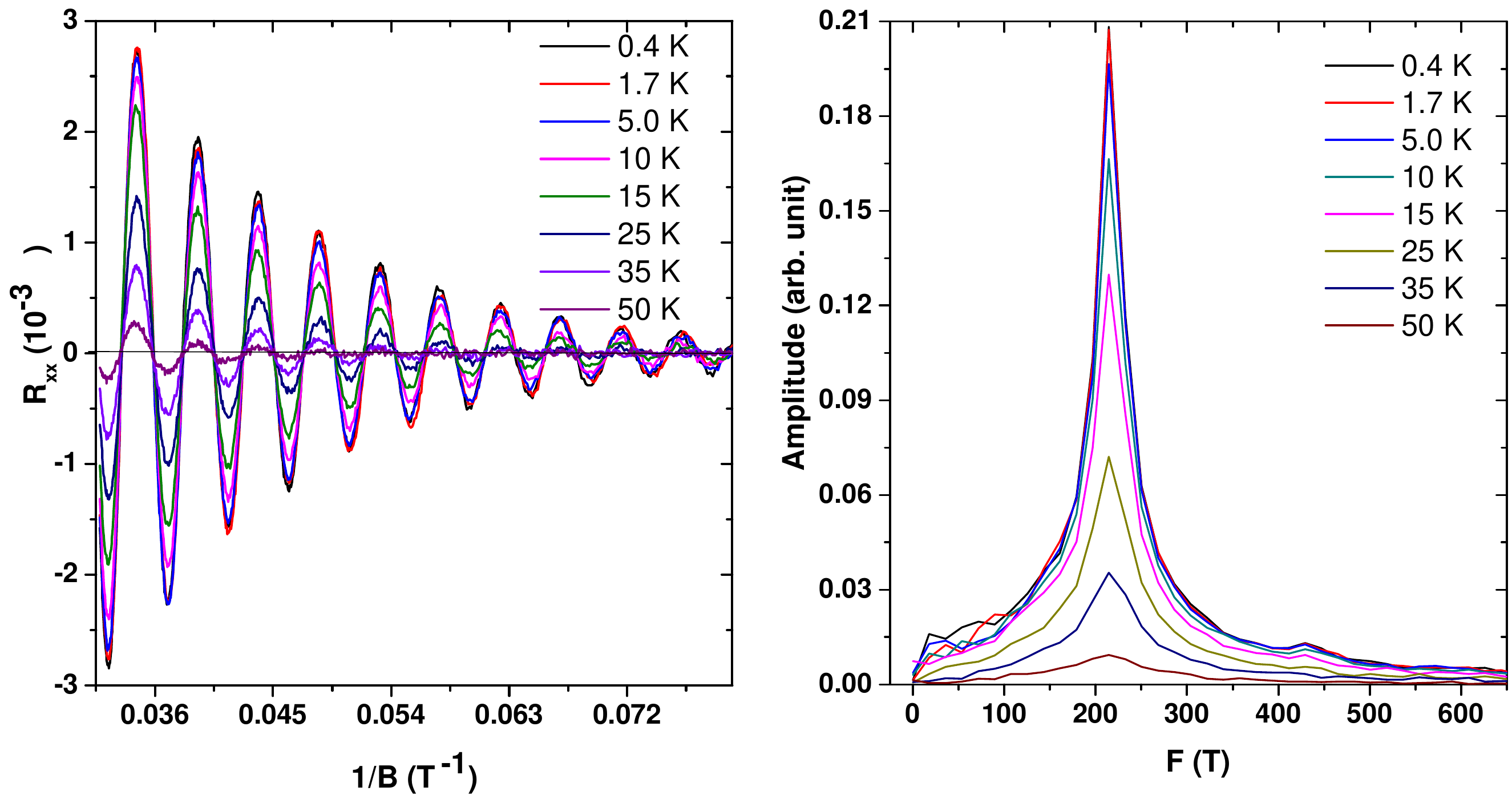}
  \caption{(a) Temperature dependence of quantum oscillations of a Sb$_2$Te$_2$Se single crystal measured at $\theta$=0$^{o}$, obtained after subtracting a smooth polynomial background. (b) Fast Fourier transform (FFT) of quantum oscillations in figure 3(a).}\label{FFT}
\end{figure}
\begin{figure}
  \centering
  \includegraphics[width=1.0\linewidth]{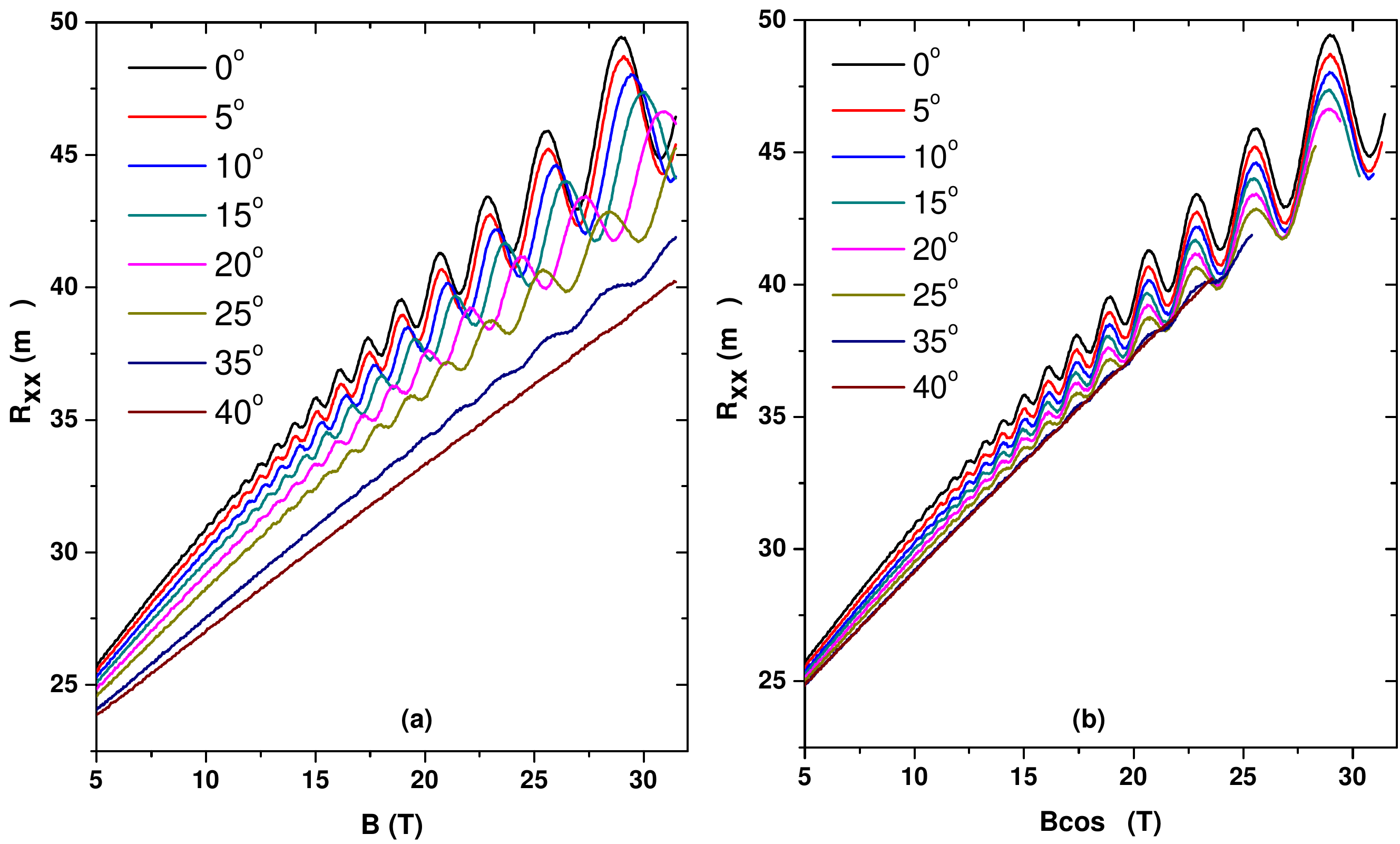}
  \caption{(a) SdH oscillations of a Sb$_2$Te$_2$Se single crystal at different tilt angles, $\theta$. The oscillations cannot be distinguished above $\theta$=40$^{o}$. (b) SdH oscillations as a function of the normal component of the applied field, $B$cos$\theta$.}\label{Angle}
\end{figure}

\indent In order to provide further evidence for the surface states origin of the quantum oscillations, we have calculated the Berry phase, $\beta$, from the Landau level (LL) fan diagram, which should be $\beta$=0.5 for Dirac particles and 0 for normal fermions. Figure [5(a)] shows magnetoconductivity $\sigma_{xx}(B)$, obtained by using the formula $\sigma_{xx}(B)=\rho_{xx}/(\rho_{xx}^{2}+\rho_{xy}^{2})$, where $\rho_{xx}$ and $\rho_{xy}$ are longitudinal and Hall resistivity, respectively. We let 1/B$_{max}$ and 1/B$_{min}$ represent the positions of maxima and minima of the quantum oscillations in $\sigma_{xx}(B)$, respectively. We subtracted a linear drift term in the quantum oscillations and the positions of minima and maxima were assigned integer and half integer values respectively to construct the LL fan diagram shown in figure [5(b)]. We have extrapolated 1/B to zero for both maxima and minima separately in the LL fan diagram. In the limit 1/B$\rightarrow$0, we have obtained $\beta$=0.45$\pm$0.02 and 0.42$\pm$0.02 for minima and maxima extrapolations, respectively. These $\beta$ values are in agreement with each other and are very close to the theoretical value 0.5 for the Dirac particles. This further confirms that quantum oscillations originate from topological surface states.\\
\begin{figure}
  \centering
  \includegraphics[width=1.0\linewidth]{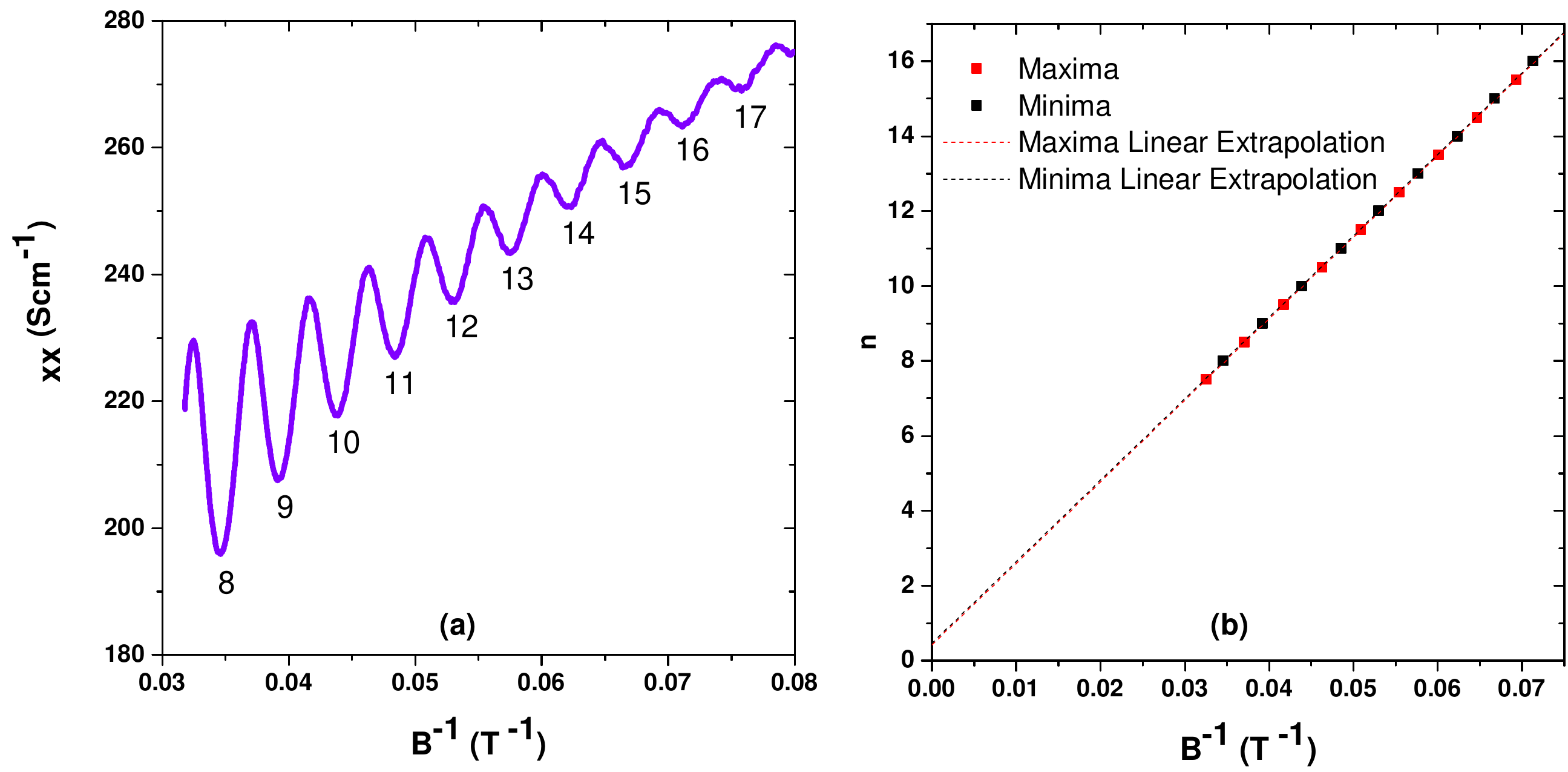}
  \caption{(a) Magnetoconductivity of a Sb$_2$Te$_2$Se single crystal plotted as a function of the inverse of magnetic field B$^{-1}$. The minima positions are assigned as integer Landau level indices. (b) Landau level fan diagram of a Sb$_2$Te$_2$Se single crystal. The positions of minima and maxima in magnetoconductivity are shown by red and black dots, respectively. The solid blue line is the linear extrapolation of the linear fit to the data.}\label{Angle}
\end{figure}
\begin{figure}
  \centering
  \includegraphics[width=1.0\linewidth]{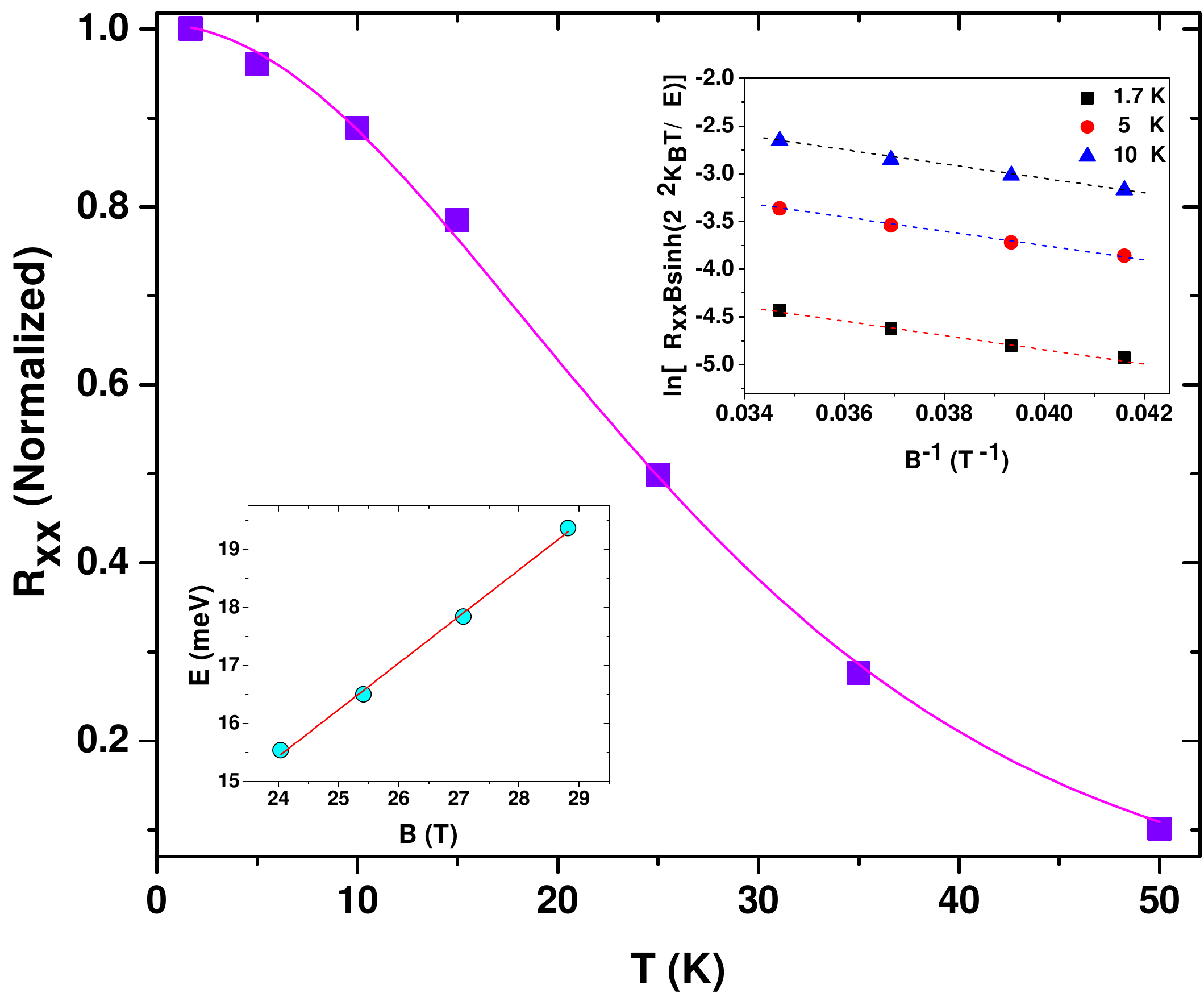}
  \caption{Temperature-dependence of the amplitude of SdH oscillations ($\Delta R_{xx}$) at B=28.8 T. The red line represents the fit using LK equation\cite{Shoenberg:22}. Upper right inset: Dingle plot to determine the Dingle temperature $T_D$ and the carrier lifetime $\tau$. The dotted lines are linear fitting to the data. Lower left inset: Landau level energy spacing ($\Delta E$) vs B plot to estimate effective cyclotron mass $m_{cyc}$. The solid red line is the linear fit.}\label{LK}
\end{figure}
\indent The frequency of quantum oscillations at $\theta$=0$^{o}$ is $f$=215 T. Using Onsagar's relation $f=\hbar/(2e)k_{F}^2$, the Fermi momentum corresponds to $k_F$=8$\times$10$^6$ cm$^{-1}$. Other physical parameters characterizing the surface states were calculated using Lifshitz Kosevich (LK) theory\cite{Shoenberg:22}. From LK analyses, as shown in figure [6] and lower left inset, the cyclotron mass is estimated to be $m_{cyc}$=0.1$m_0$, where $m_0$ is the rest mass of an electron. Using the linear dispersion relation for the surface state
$v_F=\hbar k_F/m_{cyc}$, we have estimated the Fermi velocity $v_F$=6.7$\times$10$^{5}$ ms$^{-1}$. Following the standard Dingle temperature analyses (upper inset in figure [6]), we have calculated the Dingle temperature, $T_D$=35 K. With the value of $T_D$=35 K, the surface carrier life time $\tau=\hbar/2\pi K_BT_D$ is estimated to be $\tau$=3.5$\times$10$^{-14}$ s. Similarly, other physical parameters like the mean free path $l=v_F\tau$ , mobility $\mu=e\tau/m_{cyc}$, and Fermi energy $E_F$ are estimated to be 23 nm, 600 $cm^2/V s$, and 250 meV, respectively. These physical parameters are comparable with those of the previous reports on other topological systems\cite{Ren:21}$^,$\cite{Pan:23}.

\indent From the angle dependence of quantum oscillations and the Berry phase calculations, we have proven the existence of topological surface states in the Sb$_2$Te$_2$Se single crystal. It is interesting to observe topological surface states in the metallic Sb$_2$Te$_2$Se with such a high oscillation frequency. We did not observe a second frequency as seen in our previous study on Bi$_{2}$Se$_{2.1}$Te$_{0.9}$ samples \cite{Shrestha:15}$^,$\cite{Shrestha1:16}. The higher value of the Fermi wave vector $k_{F}$ in Sb$_2$Te$_2$Se indicates the higher position of the Fermi energy from the Dirac point. Following a similar argument as in our previous work on Bi$_{2}$Se$_{2.1}$Te$_{0.9}$, although the Fermi energy is higher in the Sb$_2$Te$_2$Se crystal, it still cuts the two valence band maxima in the band structure. This is why it still shows a metallic behavior and has $p$-type bulk carriers. Due to the higher value of the bulk states Fermi wave vector, an even higher magnetic field strength is needed to observe quantum oscillations from the bulk states as constrained by the relation $f/B_{n} -\beta =(n-1)$, where $n$ represents the Landau level. This explains qualitatively the dominance of topological surface states in the Sb$_2$Te$_2$Se single crystal. The bulk states interference might be seen at higher magnetic field, beyond the 31 T maximum magnetic field of this study. The smaller value of the electron mean free path ($l$=23nm) in the Sb$_2$Te$_2$Se sample as compared to 39 nm for the Bi$_{2}$Se$_{2.1}$Te$_{0.9}$ sample indicates the presence of more scattering centers in the Sb$_2$Te$_2$Se sample. The higher RRR=$\rho_{xx}$(300K)/$\rho_{xx}$(20K)=12 for the Bi$_{2}$Se$_{2.1}$Te$_{0.9}$ sample as compared to 3.5 for the Sb$_2$Te$_2$Se sample further supports the presence of larger scattering centers in the Sb$_2$Te$_2$Se sample. The higher the number of scattering centers, the smaller the electron mobility. This is consistent with the lower value of electron mobility ($\mu$=600 $cm^2/V s$) in Sb$_2$Te$_2$Se than the 2200 $cm^2/V s$ in the Bi$_{2}$Se$_{2.1}$Te$_{0.9}$ sample.

\section*{acknowledgements}
This work is supported in part by the U.S. Air Force Office
of Scientific Research, the T. L. L. Temple Foundation, the J. J. and R. Moores Endowment, and the State of Texas through the TCSUH. V. Marinova acknowledges support from the Bulgarian Science Fund, project FNI-T-02/26. A portion of this work was performed at the National High Magnetic Field Laboratory, which is supported by National Science Foundation Cooperative Agreement No. DMR-1157490 and the State of Florida. The work at Idaho National Laboratory is supported by Department of Energy, Office of Basic Energy Sciences, Materials Sciences, and Engineering Division and through grant DOE FG02-01ER45872.

\bibliography{STS}

\end{document}